\documentclass[12pt,amssymb,amsmath,tightenlines]{article}
\usepackage{amsmath, amssymb, amsthm, latexsym, mathrsfs}
\usepackage{color}

\newcommand{\rd}{\mbox{$\rm d$}}
\newcommand{\PR}{\mathbb{P}}

\newcommand{\R}{\mathbb{R}}
\newcommand{\E}{\mathbb{E}}
\newcommand{\B}{\mathbb{B}}
\newcommand{\F}{\mathcal{F}}

\newcommand{\nn}{\nonumber}

\theoremstyle{definition}

\numberwithin{equation}{section}

\title{\bf{Pricing Fixed-Income Securities in an Information-Based Framework}}
\begin{document}

\author{Lane P. Hughston$^{\ast}$ and Andrea Macrina$^{\dag\,\ddag}$}
\date{}
\maketitle
\begin{center}
$^{\ast}$ Department of Mathematics, Imperial College London\\London SW7 2AZ, United Kingdom\\
$^{\dag}$ Department of Mathematics, King's College London\\London WC2R 2LS, United Kingdom\\
$^{\ddag}$ Institute of Economic Research, Kyoto University\\
Kyoto 606-8501, Japan
\end{center}
\begin{abstract}
In this paper we introduce a class of information-based models for
the pricing of fixed-income securities. We consider a set of
continuous-time processes that describe the flow of information
concerning market factors in a monetary economy. The nominal pricing
kernel is at any given time assumed to be given by a function of the
values of information processes at that time. By use of a
change-of-measure technique we derive explicit expressions for the
price processes of nominal discount bonds, and deduce the associated
dynamics of the short rate of interest and the market price of risk.
The interest rate positivity condition is expressed as a
differential inequality. We proceed to model the price level, which
at any given time is also taken to be a function of the values of
the information processes at that time. A simple model for a
stochastic monetary economy is introduced in which the prices of
nominal discount bonds and inflation-linked notes can be expressed
in terms of aggregate consumption and the liquidity benefit
generated by the money supply.
\\
This version: {\today}.
\\ Email: lane.hughston@imperial.ac.uk, andrea.macrina@kcl.ac.uk\\
\end{abstract}

\section{Introduction} \label{sec:1}
The idea of information-based asset pricing (Macrina 2006, Brody
{\it et al.} 2007, 2008a,b, Hughston \& Macrina 2008) is that the
market filtration should be modeled in such a way that it is
generated by a set of processes that carry information about the
future cash flows generated by tradable securities. One can regard
each such cash flow as a random variable that is in turn given by a
function of one or more independent random variables called ``market
factors" or, more succinctly, ``$X$-factors". The information
processes that generate the market filtration are associated with
the various $X$-factors in such a way that the value of each
$X$-factor is revealed at some designated time by the associated
information process. The simplest examples of information processes
are those based on Brownian bridges (Brody {\it et al.} 2007, 2008a,
Rutkowski \& Yu 2007), and gamma bridges (Brody {\it et al.} 2008b),
which lead to highly tractable asset pricing models; more general
information processes can be constructed based on L\'evy random
bridges (Hoyle {\it et al.} 2009).

The purpose of the present paper is to present a simple class of
information-based models for interest rates, foreign exchange, and
inflation.  The point of view is the following. We retain the
premise that the $X$-factors represent the fundamental factors, the
values of which are revealed from time to time, that determine the
cash flows generated by primary securities. We also accept the view
that the market filtration is generated collectively by the
information processes associated with these factors. In a
macroeconomic setting with a dynamic equilibrium, it is appropriate
to assume the existence of a universal pricing kernel associated
with the choice of a suitable base currency. We shall call this the
nominal pricing kernel associated with the given base currency. The
pricing kernel is necessarily adapted to the market filtration, and
is therefore given by a functional of the trajectories of the
information processes up to the time at which the value of the
pricing kernel is to be determined. A similar property holds for the
pricing kernel associated with any other currency or unit of
exchange. The models for interest rates and foreign exchange that we
develop are characterized by the following additional assumptions:
(a) that the information processes collectively have the Markov
property with respect to the market filtration, and (b) that the
pricing kernels associated with each currency under consideration
can at any given time be expressed as a function of the values taken
by the information processes at that time. In the case of inflation,
we take a similar point of view, adapting the ``foreign exchange
analogy'' (Hughston 1998, Jarrow \& Yildirim 2003, Mercurio 2005,
Brody {\it et al.} 2008, Hinnerich 2008). In this scheme the price
level is given by the ratio of the real and the nominal pricing
kernels. These are given, in the models developed in the present
paper, by functions of the current levels of the relevant
information processes.

\section{One-factor models}

For simplicity we consider first the case of a single $X$-factor and
a single information process. The resulting theory can be worked out
rather explicitly, and from this example one can see how the general
case can be approached when there are several currencies and many
$X$-factors. In the single-factor case we proceed as follows.

The market will be modelled as usual by a probability space
$(\Omega,\F,\PR)$ equipped with a filtration $\{\F_t\}_{t\ge 0}$. We
assume that $\PR$ is  the ``real" probability measure, and that
$\{\F_t\}$ is the market filtration. The filtration will be modelled
in the following manner. Let time $0$ denote the present, and fix a
time $U>0$. We introduce a continuous random variable $X_{U}$ taking
values in $\R$, with probability density $p(x)$. The restriction to
a continuous random variable is for convenience. With this
``$X$-factor'' we associate an information process
$\{\xi_{tU}\}_{0\le t\le U}$ defined by
\begin{equation}
\xi_{tU}=\sigma t X_{U}+\beta_{tU}.
\end{equation}

Here $\sigma$ is an information flow-rate parameter, and the
Brownian bridge process $\{\beta_{tU}\}_{0\le t\le U}$ is taken to
be independent of the market factor $X_{U}$. As remarked in Brody
{\it et al.} 2007 (see also Rutkowski \& Yu 2007) it is a
straightforward exercise making use of well-known properties of the
Brownian bridge to show that $\{\xi_{tU}\}$ has the Markov property
with respect to its own filtration. We shall assume that
$\{\xi_{tU}\}$ generates the market filtration. Hence for each
$t\in[0,U]$ the sigma-algebra $\F_t$ is defined by
\begin{equation}
\F_t=\sigma\left(\{\xi_{sU}\}_{0\le s\le t}\right).
\end{equation}

It should be evident that $U$ acts as a kind of ``sunset'' for the
economy, that there is only one piece of information to be revealed,
and once it has been revealed then that is the end of the story.
This is of course an artifact of the simplicity of our assumptions,
and in a more realistic model we can expect the revelation of
$X$-factors to proceed indefinitely into the future, the more
distant ones being generally less important than the nearer.

The pricing kernel $\{\pi_{t}\}$ will be assumed to be given by a
positive function of time $t$ and the value of the information
process at $t$. Thus, we have
\begin{equation}\label{simplePK}
\pi_t=F(t,\xi_{tU}).
\end{equation}

Given the pricing kernel, we can work out the value processes of
various assets. In the simple economy under consideration, the
``primary'' assets are those that deliver a single cash flow at time
$U$ given by a suitably integrable function $H(X_U)$ that depends on
the outcome $X_U$. The value of such a security at time $t\le U$ is
given by
\begin{equation}
H_t=\frac{1}{\pi_t}\E^{\PR}\left[\pi_U H(X_U)\,\vert\,\F_t\right].
\end{equation}
For each choice of $H(X_U)$ we obtain a tradable security. We also
consider the discount-bond system associated with the given pricing
kernel. Let us write $P_{tT}$ for the price at time $t$ of a bond
that pays one unit of currency at time $T$ for $t\le T\le U$. Then
for each $T\in[0,U]$ we have:
\begin{equation}\label{basic bond}
P_{tT}=\frac{1}{\pi_t}\E^{\PR}\left[\pi_T\,\vert\,\F_t\right].
\end{equation}
Finally, we consider various ``derivative'' assets. These deliver
prescribed cash flows at one or more times in the interval $(0,U)$
determined by the values of the basic assets and the discount bonds
at various times. More generally we can consider ``information
derivatives'' for which the cash flows can depend in an essentially
arbitrary way on the information available up to the time of the
cash flow. For example, let the payoff of a security at time $T$ be
given by $G(T,\xi_{TU})$ where $G(t,\xi)$ is a function of two
variables. Then the value of this asset at $t\le T$ is
\begin{equation}
G_t=\frac{1}{\pi_t}\E^{\PR}\left[\pi_T\,
G(T,\xi_{TU})\,\vert\,\F_t\right].
\end{equation}
We observe that the value at time $t$ of a $T$-maturity option on a
primary security takes this form.
\section{Interest rates in a one-factor model}
Let us consider in more detail the properties of discount bonds.
Recalling that the information process has the Markov property, we
see that (\ref{basic bond}) reduces to the following expression:
\begin{equation}
P_{tT}=\frac{\E^{\PR}\left[F(T,\xi_{TU})\,\vert\,\xi_{tU}\right]}{F(t,\xi_{tU})}.
\end{equation}
We proceed to work out the conditional expectation. To this end we
recall one further property of the information process. This is the
existence of the so-called ``bridge measure'' $\B$. Under the bridge
measure (Brody {\it et al.} 2007) the information process
$\{\xi_{tU}\}$ is a Brownian bridge over the interval $[0,U)$. The
change-of-measure density martingale for the transformation from
$\PR$ to $\B$ is given by a process $\{M_t\}_{0\le t<U}$ defined by
\begin{equation}\label{Mt}
M_t=\left(\int^{\infty}_{-\infty}p(x)\exp\left[\frac{U}{U-t}\left(\sigma
x\,\xi_{tU}-\tfrac{1}{2}\,\sigma^2 x^2 t\right)\right]\rd
x\right)^{-1}.
\end{equation}
Applying Ito's formula, one can show that
\begin{equation}\label{Mt dyn}
\frac{\rd M_t}{M_t}=-\frac{\sigma
U}{U-t}\,\E^{\PR}\left[X_U\,\vert\,\xi_{tU}\right]\rd W_t,
\end{equation}
where the process $\{W_t\}$ defined by
\begin{equation}\label{BM}
W_t=\xi_{tU}+\int^t_0\frac{1}{U-s}\,\xi_{sU}\,\rd s-\sigma
U\int^t_0\frac{1}{U-s}\,\E^{\PR}\left[X_{U}\,\vert\,\xi_{sU}\right]\rd
s.
\end{equation}
is an $(\{\F_t\},\PR)$ Brownian motion on $[0,U)$. Thus, in the
information-based approach the Brownian motions that drive asset
prices always arise as ``secondary'' objects---innovation
processes---rather than as primary drivers. For further details of
the properties of the change-of-measure martingale $\{M_t\}$ and
related processes appearing in its definition, see Macrina 2006,
chapter 3. Bearing in mind that the random variable $M_t$ can be
expressed as a function of $t$ and $\xi_{tU}$, as given by
(\ref{Mt}), we can without loss of generality introduce a positive
function $f(t,\xi_{tU})$ such that
\begin{equation}\label{Mpkernel}
\pi_t=M_t\,f(t,\xi_{tU}),
\end{equation}
and as a consequence we obtain
\begin{equation}
P_{tT}=\frac{\E^{\PR}\left[M_T\,f(T,\xi_{TU})\,\vert\,\xi_{tU}\right]}{M_t\,f(t,\xi_{tU})}.
\end{equation}
The appearance of the change-of-measure density in this formula
enables us to use the conditional version of Bayes formula to
re-express $\{P_{tT}\}$ in terms of an expectation with respect to
the bridge measure:
\begin{equation}\label{B-Bond Price}
P_{tT}=\frac{\E^{\B}\left[f(T,\xi_{TU})\,\vert\,\xi_{tU}\right]}{f(t,\xi_{tU})}.
\end{equation}
Since the information process is a $\B$-Brownian bridge we know that
at each time $t$ the random variable $\xi_{tU}$ is $\B$-Gaussian.
Armed with this fact, we proceed as follows. We introduce a random
variable $Y_{tT}$ defined by
\begin{equation}
Y_{tT}=\xi_{TU}-\frac{U-T}{U-t}\,\xi_{tU}.
\end{equation}
It is evident that $Y_{tT}$ is $\B$-Gaussian, and a short
calculation making use of properties of the Brownian bridge shows
that $Y_{tT}$ has mean zero and variance
\begin{equation}
\text{Var}^{\B}\left[Y_{tT}\right]=\frac{(T-t)(U-T)}{U-t}.
\end{equation}
We observe that $Y_{tT}$ and $\xi_{tU}$ are $\B$-independent. This
can be checked by calculating the relevant covariance. Next we
express $Y_{tT}$ in terms of a standard Gaussian variable $Y$, with
mean zero and variance unity. Thus we write
\begin{equation}
Y_{tT}=\nu_{tT}Y,
\end{equation}
where
\begin{equation}
\nu_{tT}=\sqrt{\frac{(T-t)(U-T)}{U-t}}.
\end{equation}
Then we rewrite (\ref{B-Bond Price}) in terms of $Y$ to obtain
\begin{equation}\label{Y-Bond Price}
P_{tT}=\frac{1}{f(t,\xi_{tU})}\,\E^{\B}\left[f\left(T,\nu_{tT}Y+\frac{U-T}{U-t}\,\xi_{tU}\right)\bigg\vert\,\xi_{tU}\right].
\end{equation}
Since $Y$ and $\xi_{tU}$ are $\B$-independent, the conditional
expectation in (\ref{Y-Bond Price}) reduces to a Gaussian integral
over the range of $Y$, and thus:
\begin{equation}
P_{tT}=\frac{1}{\sqrt{2\pi}\,f(t,\xi_{tU})}\int^{\infty}_{-\infty}
f\left(T,\nu_{tT}\,y+\frac{U-T}{U-t}\,\xi_{tU}\right)\,\exp\left(-\tfrac{1}{2}\,y^2\right)\rd
y.
\end{equation}
We can write this equation more compactly by introducing a function
of three variables $\tilde{f}(t,T,\xi)$ as follows:
\begin{equation}
\tilde{f}(t,T,\xi)=\frac{1}{\sqrt{2\pi}}\int^{\infty}_{-\infty}f\left(T,\nu_{tT}\,y+\frac{U-T}{U-t}\,\xi\right)\exp\left(-\tfrac{1}{2}\,y^2\right)\rd
y.
\end{equation}
Then, the bond price is given by
\begin{equation}
P_{tT}=\frac{\tilde{f}(t,T,\xi_{{tU}})}{f(t,\xi_{tU})}.
\end{equation}
For any particular choice of $f$ it is straightforward to simulate
the dynamics of the bond price since, conditional on the outcome of
the underlying factor $X_{U}$, the information process is
$\PR$-Gaussian.
\section{Pricing kernel, nominal interest rate, and market price of risk}
Let us proceed to derive the dynamics of the pricing kernel
(\ref{Mpkernel}). We apply the product rule to obtain
\begin{equation}
\rd\pi_t=f_t\,\rd M_t+M_t\,\rd f_t+\rd M_t\,\rd f_t,
\end{equation}
where $f_t=f(t,\xi_{tU})$. The dynamical equation for the
change-of-measure density martingale is (\ref{Mt dyn}). We shall
assume that $f(t,\xi)$ has a continuous first derivative with
respect to $t$, denoted $\dot{f}(t,\xi)$, and a continuous second
derivative with respect to $\xi$, denoted $f''(t,\xi)$. Hence
\begin{equation}
\rd f_t=\dot{f}_t\,\rd
t+f'_t\,\rd\xi_{tU}+\tfrac{1}{2}\,f''_t\,(\rd\xi_{tU})^2.
\end{equation}
In terms of the innovation process defined by (\ref{BM}), the
dynamical equation for $\{\xi_{tU}\}$ is
\begin{equation}\label{xi}
\rd\xi_{tU}=\frac{1}{U-t}\left(\sigma\,
U\,\E\left[X_{U}\,\vert\,\xi_{tU}\right]-\xi_{tU}\right)\rd t+\rd
W_t.
\end{equation}
Thus $(\rd\xi_{tU})^2=\rd t$, and with (\ref{Mt}) and (\ref{xi}) at
hand a calculation shows that
\begin{align}\label{dyn pi}
&\rd\pi_t=\nn\\
&M_t\left(\dot{f}_t-\frac{\xi_{tU}}{U-t}f'_t+\tfrac{1}{2}f''_t\right)\rd
t\,+\,M_t\left(f'_t-\frac{\sigma
U}{U-t}\,\E\left[X_{U}\,\vert\,\xi_{tU}\right]\,f_t\right) \rd W_t.
\end{align}
or equivalently,
\begin{align}
\frac{\rd\pi_t}{\pi_t}=\frac{1}{f_t}&\left(\dot{f}_t-\frac{\xi_{tU}}{U-t}\,f'_t+\tfrac{1}{2}f''_t\right)\rd
t\nn\\
&+\,\frac{1}{f_t}\left(f'_t-\frac{\sigma
U}{U-t}\,\E\left[X_{U}\,\vert\,\xi_{tU}\right]\, f_t\right)\rd W_t.
\end{align}
The drift of the pricing kernel is minus the short rate of interest,
and the volatility is minus the market price of risk:
\begin{equation}\label{NPKD}
\frac{\rd\pi_t}{\pi_t}=-r_t\,\rd t-\lambda_t\,\rd W_t.
\end{equation}
Comparing coefficients, we deduce that the nominal short rate and
market price of risk are given respectively by
\begin{equation}
r_t=\frac{1}{f_t}\left(\frac{\xi_{tU}}{U-t}\,f'_t-\tfrac{1}{2}\,f''_t-\dot{f}_t\right),
\end{equation}
and
\begin{equation}
\lambda_t=\frac{\sigma
U}{U-t}\,\E\left[X_{U}\,\vert\,\xi_{tU}\right]-\frac{f'_t}{f_t}.
\end{equation}
It is natural in the context of some applications to impose the
condition that the short rate should be positive. This condition is
evidently given by
\begin{equation}\label{SPDE ineq}
\frac{\xi_{tU}}{U-t}\,f'_t-\tfrac{1}{2}\,f''_t-\dot{f}_t>0.
\end{equation}
The interest-rate positivity condition is equivalent to the
following differential inequality:
\begin{equation}
\frac{x}{U-t}\,f'(t,x)-\tfrac{1}{2}f''(t,x)-\dot{f}(t,x)>0.
\end{equation}
\section{Pricing in a multi-factor setting}
We introduce a set of $X$-factors $\{X_{T_1},\ldots,X_{T_n}\}$,
labeled by a series of dates $T_k$ $(k=1,\ldots,n)$ such that
$0<T_1<\ldots<T_n$. With each $X$-factor we associate an information
process $\{\xi_{tT_k}\}$ defined by
\begin{equation}
\xi_{tT_k}=\sigma_k\,t\,X_{T_k}+\beta_{tT_k}.
\end{equation}
The information processes associated with different $X$-factors are
taken to be independent. The market filtration $\{\F_t\}$ is assumed
to be generated by collection of information processes:
\begin{equation}
\F_t=\sigma\left(\left\{\xi_{sT_1}\right\}_{0\le s\le
t},\ldots,\left\{\xi_{sT_n}\right\}_{0\le s\le t}\right).
\end{equation}
As a generalisation of the Markov model introduced in the previous
section, we consider the following multi-factor model for $\pi_t$:
\begin{equation}\label{multiMpkernel}
\pi_t=M^{(1)}_{t}\cdots
M^{(n)}_{t}\,f(t,\xi_{tT_1},\ldots,\xi_{sT_n}).
\end{equation}
Here $f(t,\xi_1,\xi_2,\ldots,\xi_n)$ is a function of $n+1$
variables. The unit-initialized $\PR$-martingales
$\{M^{(k)}_{t}\}_{k=1,\ldots,n}$ are defined by
\begin{equation}\label{PB-Mart}
\frac{\rd
M^{(k)}_t}{M^{(k)}_t}=-\frac{\sigma_k\,T_k}{T_k-t}\,\E\left[X_{T_k}\,\vert\,\xi_{tT_k}\right]\rd
W^{(k)}_t,
\end{equation}
where for each $k$ the $\PR$-Brownian motion $\{W_t^{(k)}\}$ is
defined by
\begin{equation}
W^{(k)}_t=\xi_{tT_k}+\int^t_0\frac{1}{T_k-s}\,\xi_{sT_k}\rd
s-\sigma_k
T_k\int^t_0\frac{1}{T_k-s}\,\E\left[X_{T_k}\,\vert\,\xi_{sT_k}\right]\rd
s.
\end{equation}
Since the information processes are independent, it follows that
\begin{equation}
\rd W^{(j)}\rd W^{(k)}=\delta^{jk}\,\rd t.
\end{equation}
Let us focus on the pricing of a nominal discount bond with maturity
$T<T_1$. The price of the bond is given by:
\begin{equation}
P_{tT}=\frac{\E^{\PR}\left[M^{(1)}_{T}\cdots
M^{(n)}_{T}\,f\left(T,\xi_{TT_1},\ldots,\xi_{TT_n}\right)
\,\big\vert\,\xi_{tT_1},\ldots,\xi_{tT_n}\right]} {M^{(1)}_{t}\cdots
M^{(n)}_{t}\,f\left(t,\xi_{tT_1},\ldots,\xi_{tT_n}\right)}.
\end{equation}
Here we have used the fact that the information processes are
Markovian. Since the information processes are independent, the
product of $\PR$-martingales given by $M^{(1)}_{t}\cdots
M^{(n)}_{t}$ for $t$ in the interval $[0,T_1)$ is itself a
$\PR$-martingale. This martingale induces a bridge measure that has
the effect of turning the information processes into $\B$-Brownian
bridges. More precisely, under the bridge measure each information
process has, over the interval $[0,T_1)$, the distribution of a
standard Brownian bridge on the interval from $0$ to the termination
time of the information process. Thus we have
\begin{equation}
P_{tT}=\frac{\E^{\B}\left[f\left(T,\xi_{TT_1},\ldots,\xi_{TT_n}\right)
\,\big\vert\,\xi_{tT_1},\ldots,\xi_{tT_n}\right]}
{f\left(t,\xi_{tT_1},\ldots,\xi_{tT_n}\right)},
\end{equation}
where all of the variables appearing are $\B$-Gaussian. Next we
introduce a set of random variables
$Y^{(1)}_{tT},Y^{(2)}_{tT},\ldots,Y^{(n)}_{tT}$ defined by
\begin{equation}\label{substitution 1}
Y^{(k)}_{tT}=\xi_{TT_k}-\frac{T_k-T}{T_k-t}\,\xi_{tT_k}.
\end{equation}
Since $\{\xi_{tT_k}\}$ is a $\B$-Brownian bridge, it follows that
$Y^{(k)}_{tT}$ is Gaussian with mean zero and variance
\begin{equation}
\textrm{Var}\,^{\B}\left[Y^{(k)}_{tT}\right]=\frac{(T-t)(T_k-T)}{T_k-t}.
\end{equation}
We introduce therefore an $n$-dimensional set of standard Gaussian
variables $\left(Y_1,\ldots,Y_n\right)$. The variable $Y_k$ stands
in relation to $Y^{(k)}_{tT}$ via the formula
\begin{equation}\label{substitution 2}
Y^{(k)}_{tT}=\nu^{(k)}_{tT}\,Y_k,
\end{equation}
where
\begin{equation}
\nu^{(k)}_{tT}=\sqrt{(T-t)(T_k-T)/(T_k-t)}.
\end{equation}

In terms of the standard Gaussian random variables, the bond price
at time $t$ can thus be written in the form
\begin{align}
&P_{tT}=\nn\\
&\frac{\E^{\B}\left[f\left(T,\,
\nu^{(1)}_{tT}Y_1+\frac{T_1-T}{T_1-t}\xi_{tT_1}\,,\, \ldots \, ,
\nu^{(n)}_{tT}Y_n+\frac{T_n-T}{T_n-t}\xi_{tT_n}\right)\,\vert\,\xi_{tT_1}
\cdots \xi_{tT_n}\right]}
{f\left(t,\xi_{tT_1},\ldots,\xi_{tT_n}\right)}.
\end{align}
We observe that the random variables $Y^{(k)}_{tT}$ and $\xi_{tT_k}$
are $\B$-independent. The expression for the bond price thus reduces
to a definite integral:
\begin{eqnarray}
&&P_{tT}=\int^{\infty}_{-\infty}\cdots\int^{\infty}_{-\infty}\frac{f(T,\,\nu^{(1)}_{tT}\,y_1+\frac{T_1-T}{T_1-t}\,\xi_{tT_1}\,,\,\ldots\,,\,\nu^{(n)}_{tT}\,y_n+\frac{T_n-T}{T_n-t}\,\xi_{tT_n})}
{f(t,\xi_{tT_1},\ldots,\xi_{tT_n})}\nn\\
&&\hspace{3.5cm}\times\frac{1}{\left(\sqrt{2\pi}\right)^n}\exp\left[-\tfrac{1}{2}\left(y^2_n+\ldots+y^2_1\right)\right]\rd
y_1\cdots\rd y_n.\nn\\
\end{eqnarray}
That is to say, we obtain an expression of the form
\begin{equation}
P_{tT}=\frac{\tilde{f}\left(t,T,\xi_{tT_1},\ldots,\xi_{tT_n}\right)}
{f\left(t,\xi_{tT_1},\ldots,\xi_{tT_n}\right)},
\end{equation}
where
\begin{align}
&\tilde{f}(t,T,\xi_1,\xi_2,\ldots,\xi_n)\nn\\
&=\int^{\infty}_{-\infty}\cdots\int^{\infty}_{-\infty}
f\left(T,\,\nu^{(1)}_{tT}\,y_1+\frac{T_1-T}{T-t}\,\xi_{1}\,,\,\ldots\, ,\,\nu^{(n)}_{tT}\,y_n+\frac{T_n-T}{T-t}\,\xi_{n}\right)\nn\\
&\hspace{2.75cm}\times\frac{1}{\left(\sqrt{2\pi}\right)^n}\,\exp\left[-\tfrac{1}{2}\left(y^2_n+\ldots+y_1^n\right)\right]\rd
y_n\cdots y_1.
\end{align}
A multi-currency environment, with an interest rate system for each
currency, can be handled similarly. We consider $N+1$ currencies,
writing $\{\pi_t\}$ for the pricing kernel of the ``domestic'' or
``base'' currency, and $\{\pi_t^i\}$, $i=1,\ldots,N$, for the
pricing kernels of the foreign currencies. We introduce $n$
information processes, and assume that each of the pricing kernels
is a function of the current levels of the information processes.
The prices associated with the $N$ foreign currencies, expressed in
units of the domestic currency, are the ratios of the various
foreign pricing kernels to the domestic pricing kernel. For a
realistic model, we expect to have $n\ge 2N+1$.
\section{Positivity condition}
Let us consider the class of functions $f(t,\xi_1,\ldots,\xi_n)$ for
which the pricing kernel (\ref{multiMpkernel}) is a supermartingale.
We need thus to derive the dynamics of the pricing kernel and to
work out its drift. Let us assume that $f(t,\xi_1,\ldots,\xi_n)$ is
in $C^{1,2}(\R^+\times\R^n)$. We let $\dot{f}$ denote the derivative
with respect to $t$, $\partial_k f$ the derivative with respect to
the $k$-th coordinate, and $\partial_{kk} f$ the second derivative
with respect to the $k$-th coordinate. Then the dynamical equation
of the pricing kernel is:
\begin{eqnarray}
\frac{\rd\pi_t}{\pi_t}&=&\frac{1}{f_t}\left[\dot{f}+\sum^n_{k=1}\left(\tfrac{1}{2}\,\partial_{kk}
f_t-\frac{\xi_{tT_k}}{T_k-t}\,\partial_k f_t\right)\right]\rd
t\nn\\
&+&\frac{1}{f_t}\sum^n_{k=1}\left(\partial_k f_t-\frac{\sigma_k
T_k}{T_k-t}\,X_{tT_k}\,f_t\right)\rd W^{k}_t,
\end{eqnarray}
where $X_{tT_k}=\E[X_{T_k}\,\vert\,\xi_{tT_k}]$. The multi-factor
short rate process is thus
\begin{equation}
r_t=\frac{1}{f_t}\left[\sum^{n}_{k=1}\left(\frac{\xi_{tT_k}}{T_k-t}\,\partial_k
f_t-\frac{1}{2}\,\partial_{kk} f_t\right)-\dot{f_t}\right],
\end{equation}
and for the $k$-th component of the market price of risk vector we
obtain
\begin{equation}
\lambda^k_t=\frac{1}{f_t}\left(\frac{\sigma_k
T_k}{T_k-t}\,X_{tT_k}f_t-\partial_k f_t\right).
\end{equation}
If we impose the condition of a positive short rate process, then
the following condition needs to be satisfied:
\begin{equation}\label{mulivariate SPDE ineq}
\sum^{n}_{k=1}\left(\frac{\xi_{tT_k}}{T_k-t}\,\partial_k
f_t-\tfrac{1}{2}\,\partial_{kk} f_t\right)-\dot{f_t}> 0.
\end{equation}
A sufficient condition for (\ref{mulivariate SPDE ineq}) is that the
function $f(t,\xi_1,\ldots,\xi_n)$ satisfies
\begin{equation}
\sum^n_{k=1}\left[\frac{x_k}{T_k-t}\,\partial_k
f(t,\xi_1,\ldots,\xi_n)-\tfrac{1}{2}\,\partial_{kk}f(t,\xi_1,\ldots\,\xi_n)\right]-\dot{f}(t,\xi_1,\ldots,\xi_n)>0.
\end{equation}
\section{Inflation-linked products}
The technique used for nominal discount bonds can be adapted to the
pricing of inflation-linked assets. In what follows we focus on the
pricing of inflation-linked discount bonds. We denote the price
level (e.g., the consumer price index) by $\{C_t\}$, and note that
the relation between the nominal pricing kernel $\{\pi_t\}$, the
real pricing kernel $\{\pi^R_t\}$, and the price level is
\begin{equation}\label{Cpipi}
C_t=\frac{\pi^R_t}{\pi_t}.
\end{equation}
We take the view that the dynamics of the price level should be
derived from the dynamics of the pricing kernels. We return to this
point shortly, when we introduce the elements of a stochastic
monetary economy. Once models for the nominal and the real pricing
kernels have been constructed, then the dynamical equation of the
price level follows as a result of (\ref{Cpipi}).

It will be convenient to define an inflation-linked discount bond as
a security that at its maturity $T$ generates a single cash flow
equal to the price level $C_T$ prevailing at that time. Thus, the
price $\{Q_{tT}\}_{0\le t\le T}$ of an inflation-linked discount
bond is given by
\begin{equation}
Q_{tT}=\frac{\E^{\PR}\left[\pi_T\,C_T\,\vert\,\F_t\right]}{\pi_t}.
\end{equation}
Using the relationship (\ref{Cpipi}) we can write this alternatively
as
\begin{equation}
Q_{tT}=\frac{\E^{\PR}\left[\pi^R_T\,\vert\,\F_t\right]}{\pi_t}.
\end{equation}

We shall construct models for the nominal and real pricing kernels
following the approach presented in the earlier sections. For
simplicity we introduce a pair of independent market factors
$X_{T_1}$ and $X_{T_2}$, where $0\le t\le T<T_1<T_2$, along with the
associated Brownian-bridge information processes $\{\xi_{tT_1}\}$
and $\{\xi_{tT_2}\}$ that generate the filtration, and set
\begin{equation}
\pi_t=M^{(1)}_t M^{(2)}_t\,f\left(t,\xi_{tT_1},\xi_{tT_2}\right),
\end{equation}
and
\begin{equation}
\pi^R_t=M^{(1)}_t M^{(2)}_t\,g\left(t,\xi_{tT_1},\xi_{tT_2}\right),
\end{equation}
where the $\mathbb{P}$-martingales $\{M^{(1)}_t\}$ and
$\{M^{(2)}_t\}$ are given by expressions analogous to (\ref{Mt}).
The price of an inflation-linked bond is thus
\begin{equation}
Q_{tT}=\frac{\E^{\PR}\left[M^{(1)}_T
M^{(2)}_T\,g(T,\xi_{TT_1},\xi_{TT_2})\,\big\vert\,\xi_{tT_2},\xi_{tT_2}\right]}
{M^{(1)}_t M^{(2)}_t\,f\left(t,\xi_{tT_1},\xi_{tT_2}\right)}.
\end{equation}
Here we have made use of the Markov property of the information
processes. Then we change measure from $\PR$ to $\B$ to obtain
\begin{equation}
Q_{tT}=\frac{\E^{\B}\left[g(T,\xi_{TT_1},\xi_{TT_2})\,\big\vert\,\xi_{tT_2},\xi_{tT_1}\right]}
{f(t,\xi_{tT_1},\xi_{tT_2})}.
\end{equation}
The conditional expectation reduces to a Gaussian integral and the
bond price process can be expressed as follows:
\begin{eqnarray}\label{2-dim ILBond}
&&Q_{tT}=\frac{1}{2\pi}\int^{\infty}_{-\infty}\int^{\infty}_{-\infty}\frac{g(T,\nu^{(1)}_{tT}y_1+\frac{T_1-T}{T_1-t}\,\xi_{tT_1},\nu^{(2)}_{tT}y_2+\frac{T_2-T}{T_2-t}\,\xi_{tT_2})}
{f(t,\xi_{tT_1},\xi_{tT_2})}\nn\\
&&\hspace{3.5cm}\times\exp\left[-\tfrac{1}{2}\left(y^2_1+y^2_2\right)\right]\rd
y_1\rd y_2.
\end{eqnarray}
In such a setting the nominal discount bond has the following price:
\begin{eqnarray}\label{2-dim NomBond}
&&P_{tT}=\frac{1}{2\pi}\int^{\infty}_{-\infty}\int^{\infty}_{-\infty}\frac{f(T,\nu^{(1)}_{tT}y_1+\frac{T_1-T}{T_1-t}\,\xi_{tT_1},\nu^{(2)}_{tT}y_2+\frac{T_2-T}{T_2-t}\,\xi_{tT_2})}
{f(t,\xi_{tT_1},\xi_{tT_2})}\nn\\
&&\hspace{3.5cm}\times\exp\left[-\tfrac{1}{2}\left(y^2_1+y^2_2\right)\right]\rd
y_1\rd y_2.
\end{eqnarray}

The dynamics of the real pricing kernel can be computed analogously
to that of the nominal pricing kernel. Since the real interest rate
may be positive or negative, there is no supermartingale condition
on $g$. Inserting the expressions for the nominal and the real
pricing kernels in (\ref{Cpipi}), one obtains the dynamics of the
price level. In the case of a general $n$-factor model the pricing
kernels are given by expressions of the following form:
\begin{equation}
\pi_t=M^{(1)}_t\cdots
M^{(n)}_t\,f\left(t,\xi_{tT_1},\ldots,\xi_{tT_n}\right),
\end{equation}
and
\begin{equation}
\pi^R_t=M^{(1)}_t\cdots
M^{(n)}_t\,g\left(t,\xi_{tT_1},\ldots,\xi_{tT_n}\right).
\end{equation}
The price level is then given by
\begin{equation}
C_t=\frac{g\left(t,\xi_{tT_1},\ldots,\xi_{tT_n}\right)}
{f\left(t,\xi_{tT_1},\ldots,\xi_{tT_n}\right)},
\end{equation}
and for the dynamics of $\{C_t\}$ we obtain:
\begin{align}
\frac{\rd
C_t}{C_t}&=\left\{\frac{1}{f_t}\left[\sum^n_{k=1}\frac{\xi_{tT_k}}{T_k-t}\,\partial_k
f_t -\tfrac{1}{2}\sum^n_{k=1}\partial_k\partial_k
f_t-\dot{f}\right]\right.\nn\\
&\left.\hspace{.5cm}-\frac{1}{g_t}\left[\sum^n_{k=1}\frac{\xi_{tT_k}}{T_k-t}\,\partial_k
g_t-\tfrac{1}{2}\sum^n_{k=1}\partial_k\partial_k g_t-\dot{g}\right]\right.\nn\\
&\left.\hspace{.5cm}+\sum^n_{k=1}\frac{\sigma_k
T_k}{T_k-t}\,X_{tT_k}\left(\frac{1}{g_t}\partial_k
g_t-\frac{1}{f_t}\partial_k f_t\right)\right.\nn\\
&\left.\hspace{.5cm}-\frac{1}{g_t f_t}\sum^n_{k=1}\partial_k g_t\,\partial_k f_t+\frac{1}{f_t^2}\sum^n_{k=1}\left(\partial_k f_t\right)^2\right\}\rd t \nn\\
&+\sum^n_{k=1}\left(\frac{1}{g_t}\,\partial_k
g_t-\frac{1}{f_t}\,\partial_k f_t\right)\rd W^{(k)}_t,
\end{align}
where $X_{tT_k}=\E\left[X_{T_k}\,\vert\,\xi_{tT_k}\right]$. In this
calculation we have used the relationship
\begin{equation}
\rd\xi_{tT_k}=\frac{1}{T_k-t}\left(\sigma_k\,T_k\,X_{tT_k}-\xi_{tT_k}\right)\rd
t+\rd W_t^{(k)}.
\end{equation}
The drift of the price level is the instantaneous rate of inflation:
\begin{equation}
I_t=r_t-r_t^R+\lambda_t\left(\lambda_t-\lambda_t^R\right).
\end{equation}
The volatility of the price level on the other hand is given by the
difference between the nominal and the real market prices of risk.
Verification of these results is achieved by calculating the
dynamics of the nominal and the real pricing kernels. In particular,
we have (\ref{NPKD}) together with
\begin{equation}
\frac{\rd\pi^R_t}{\pi^R_t}=-r^R_t\,\rd t-\lambda^R_t\,\rd W_t.
\end{equation}
A calculation shows that
\begin{align}
&\frac{\rd\pi_t}{\pi_t}=\frac{1}{f_t}\left[\dot{f}_t-\sum^n_{k=1}\left(\frac{\xi_{tT_k}}{T_k-t}\,\partial_k\,f_t
-\tfrac{1}{2}\partial_k\partial_k\,f_t\right)\right]\rd
t\nn\\
&\hspace{4.75cm}+\frac{1}{f_t}\sum^n_{k=1}\left(\partial_k
f_t-\frac{\sigma_k T_k}{T_k-t}\,X_{tT_k}\,f_t\right)\rd W_t^{(k)}
\end{align}
and
\begin{align}
&\frac{\rd\pi_t^R}{\pi_t^R}=\frac{1}{g_t}\left[\dot{g}_t-\sum^n_{k=m}\left(\frac{\xi_{tT_k}}{T_k-t}\,\partial_k\,g_t
-\tfrac{1}{2}\partial_k\partial_k\,g_t\right)\right]\rd
t\nn\\
&\hspace{4.75cm}+\frac{1}{g_t}\sum^n_{k=m}\left(\partial_k
g_t-\frac{\sigma_k T_k}{T_k-t}\,X_{tT_k}\,g_t\right)\rd W_t^{(k)}.
\end{align}
The resulting dynamics of the price level can then be written in the
form
\begin{equation}
\frac{\rd
C_t}{C_t}=\left[r_t-r_t^R+\lambda_t(\lambda_t-\lambda_t^R)\right]\rd
t+(\lambda_t-\lambda_t^R)\,\rd W_t.
\end{equation}
\section{Stochastic monetary economy}
So far we have indicated how the pricing of fixed-income securities,
in particular the nominal and inflation-linked discount bond
systems, can be modelled in an information-based framework. We have
shown how the nominal and real pricing kernels, and hence the price
level, can be modelled in terms of information processes. It is our
goal now to consider the relationship between the two pricing
kernels, and to develop a simple macroeconomic model based on (a)
the liquidity benefit of the money supply, and (b) the rate of
consumption of goods and services. This will be carried out in the
context of the pricing theory developed in the previous sections. A
macroeconomic pricing model that suits the present investigation is
that presented in Hughston \& Macrina (2008). In this
finite-time-horizon model, agents maximize the expected utility
derived from the consumption of goods and services and from the
liquidity benefit of money supply.

There are three exogenously-specified processes that form the basis
of such an economy: (1) the real per-capita rate of consumption of
goods and services $\{k_t\}$, (2) the per-capita money supply
$\{m_t\}$, and (3) the rate of liquidity benefit $\{\eta_t\}$
provided per unit of money supply. The product $\eta_t m_t$ is the
instantaneous benefit rate in nominal units provided by the money
supply at time $t$. The goal of the representative agent is to find
the optimal strategy for consuming goods and services and for taking
advantage of the benefit of the supply of money. Since the liquidity
benefit is measured in nominal units, we use the price level to
convert its units to those of good and services. The ``real''
liquidity benefit rate $\{j_t\}$ is thus given by
\begin{equation}
j_t=\frac{\eta_t m_t}{C_t}.
\end{equation}

The agent's rate of utility derived from real consumption and real
liquidity benefit is modelled by a bivariate utility function
$U(x,y):\R^+\times\R^+\rightarrow\R$ of the Sidrauski (1967) type
satisfying $U_x>0$, $U_{xx}<0$, $U_y>0$, $U_{yy}<0$ and
$U_{xx}U_{yy}>(U_{xy})^2$. The strategy that delivers the agent the
highest level of total expected utility over the period $[0,T]$ is
found by maximising
\begin{equation}
I=\E\left[\int^T_0\Gamma_t U(j_t,k_t)\rd t\right],
\end{equation}
when the agent has a budget $H_0$ given by
\begin{equation}
H_0=\E\left[\int^T_0\pi_t C_t\left(j_t+k_t\right)\rd t\right].
\end{equation}
Here $\Gamma_t$ is a psychological discount factor which we take to
be deterministic. In equilibrium, a relation is thus determined
between $C_t$, $j_t$, and $k_t$.

An exact solution can be found in the case where the utility
function is of a separable bivariate logarithmic type:
\begin{equation}
U(x,y)=a\,\ln(x)+b\,\ln(y),
\end{equation}
where $a$ and $b$ are constants. We obtain the following expressions
for the pricing kernels:
\begin{align}\label{SME Kernels}
\pi_t=\frac{a_t}{\eta_t\, m_t}\quad\textrm{and}\quad
\pi_t^R=\frac{b_t}{k_t},
\end{align}
where $a_t=a\,\Gamma_t/\lambda$ and $b_t=b\,\Gamma_t/\lambda$, and
$\lambda$ is determined by the budget constraint. The price level is
then given by
\begin{equation}
C_t=\frac{b}{a}\,\frac{\eta_t m_t}{k_t}.
\end{equation}

Next we establish a link between this model and the
information-based approach presented in the previous sections. We
revert to a low-dimensional example in which the economy is driven
by a pair of macroeconomic factors $X_{T_1}$ and $X_{T_2}$ with
associated information processes $\{\xi_{tT_k}\}_{k=1,2}$. The
nominal and real pricing kernels are taken to be
\begin{align}
\pi_t=M^{(1)}_t M^{(2)}_t\,f(t,\xi_{tT_1},\xi_{tT_2})&\
&\textrm{and}&\ &\pi_t^R=M^{(1)}_t
M^{(2)}_t\,g(t,\xi_{tT_1},\xi_{tT_2}),
\end{align}
where $\{M^{(1)}_t\}$ and $\{M^{(2)}_t\}$ are the $\PR$-martingales
defined by (\ref{PB-Mart}). We assume that the real rate of
consumption, the money supply and the nominal rate of specific
liquidity benefit are given by functions of the form
\begin{align}
&k_t=k(t,\xi_{tT_1},\xi_{tT_2}),& &m_t=m(t,\xi_{tT_1},\xi_{tT_2}),&
&\eta_t=\eta(t,\xi_{tT_1},\xi_{tT_2}).&
\end{align}
By comparison with (\ref{SME Kernels}) we thus obtain the following
relationships:
\begin{equation}
f(t,\xi_{tT_1},\xi_{tT_2})=\frac{a_t}{M^{(1)}_t
M^{(2)}_t\,\eta(t,\xi_{tT_1},\xi_{tT_2})\,m(t,\xi_{tT_1},\xi_{tT_2})},
\end{equation}
and
\begin{equation}
g(t,\xi_{tT_1},\xi_{tT_2})=\frac{b_t}{M^{(1)}_t
M^{(2)}_t\,k(t,\xi_{tT_1},\xi_{tT_2})}.
\end{equation}
Applying (\ref{2-dim ILBond}) and (\ref{2-dim NomBond}), we are then
able to work out prices of bonds in a monetary economy in which
asset prices fluctuate in line with emerging information about
macroeconomic factors influencing the economy. The price of an
inflation-linked bond is:
\begin{align}
Q_{tT}&=\frac{M^{(1)}_t
M^{(2)}_t\,\eta(t,\xi_{tT_1},\xi_{tT_2})\,m(t,\xi_{tT_1},\xi_{tT_2})}{a_t}\nn\\
&\times\frac{1}{2\pi}
\int^{\infty}_{-\infty}\int^{\infty}_{-\infty}\frac{b_T\,\exp\left[-\tfrac{1}{2}\left(y_1^2+y_2^2\right)\right]\,\rd
y_1\,\rd y_2}{M^{(1)}_T\left(z(y_1)\right)
M^{(2)}_T\left(z(y_2)\right)k\left(T,z(y_1),z(y_2)\right)},
\end{align}
where
\begin{equation}
z(y_k)=\nu^{(k)}_{tT}\,y_k+\frac{T_k-T}{T_k-t}\,\xi_{tT_k},\qquad
k=1,2.
\end{equation}
The corresponding nominal discount bond price is
\begin{align}
&P_{tT}=\frac{M^{(1)}_t
M^{(2)}_t\,\eta(t,\xi_{tT_1},\xi_{tT_2})\,m(t,\xi_{tT_1},\xi_{tT_2})}{a_t}\nn\\
&\times\frac{1}{2\pi}\int^{\infty}_{-\infty}\int^{\infty}_{-\infty}\frac{a_T\,\exp\left[-\tfrac{1}{2}\left(y_1^2+y_2^2\right)\right]\,\rd
y_1\,\rd y_2}
{M^{(1)}_T\left(z(y_1)\right)M^{(2)}_T\left(z(y_2)\right)\eta\left(T,z(y_1),z(y_2)\right)m(T,z(y_1),z(y_2))}.
\end{align}
Similar formulae can be derived in the case of a separable
power-utility function. In such a situation, the nominal pricing
kernel also depends on the real rate of consumption.
\vskip 15pt \noindent {\bf Acknowledgments.} The authors are
grateful to J. Akahori, D. C. Brody, H. Gretarsson, M. Hinnerich, T.
Honda, E. Hoyle, E. Mackie, R. Miura, K. Ohashi, P. A. Parbhoo, G.
Sarais, J. Sekine, and W. T. Shaw for useful discussions. We are
grateful to seminar participants at Ritsumeikan University, Kusatsu,
Japan, the KIER-TMU International Workshop on Financial Engineering
2009, Tokyo, and ICS Hitotsubashi University, Tokyo, for helpful
comments. LPH acknowledges support from Lloyds TSB, Shell
International, and the Aspen Center for Physics.
\vskip 15pt \noindent {\bf References}.


\begin{enumerate}
\bibitem{bcl} D.~C.~Brody, J.~Crosby \& H.~Li (2008) Convexity adjustments in
inflation-linked derivatives. Risk Magazine, September 2008 issue,
124-129.

\bibitem{bhm1} D.~C.~Brody, L.~P.~Hughston \& A.~Macrina (2007) Beyond
hazard rates: a new framework for credit-risk modelling. In {\em
Advances in Mathematical Finance, Festschrift Volume in Honour of
Dilip Madan}, edited by R.~Elliott, M.~Fu, R.~Jarrow \& J.-Y.~Yen.
Birkh\"auser, Basel and Springer, Berlin.

\bibitem{bhm2} D.~C.~Brody, L.~P.~Hughston \& A.~Macrina (2008a) Information-based asset pricing.
International Journal of Theoretical and Applied Finance Vol. 11,
107-142.

\bibitem{bhm3} D.~C.~Brody, L.~P.~Hughston \& A.~Macrina (2008b) Dam rain and cumulative gain.
Proceedings of the Royal Society London A, Vol. 464, No. 2095, 1801-1822.

\bibitem{hinnerich} M.~Hinnerich (2008) Inflation-indexed swaps and
swaptions. Journal of Banking and Finance, Vol. 32, No. 11,
2293-2306.

\bibitem{hughston} L.~P.~Hughston (1998) Inflation derivatives. Merrill Lynch and
King's College London report, with added note (2004).

\bibitem{hm} L.~P.~Hughston \& A.~Macrina (2008) Information, inflation, and interest.
Advances in Mathematics of Finance. Banach Center Publications, Vol.
83, 117-138. Institute of Mathematics, Polish Academy of Science,
Warsaw.

\bibitem{jy} R.~Jarrow \& Y.~Yildirim (2003) Pricing treasury inflation
protected securities and related derivative securities using an HJM
model. Journal of Financial and Quantitative Analysis, No. 38, 409.

\bibitem{macrina} A.~Macrina (2006) An information-based framework for
asset pricing: $X$-factor theory and its applications. PhD thesis,
King's College London.

\bibitem{mercurio} F.~Mercurio (2005) Pricing inflation-indexed derivatives. Quantitative Finance, Vol. 5, No. 3, 289-302.

\bibitem{ry} M.~Rutkowski \& N.~Yu (2007) An extension of
the Brody-Hughston-Macrina approach to modeling of defaultable
bonds. International Journal of Theoretical and Applied Finance Vol.
10, 557-589.

\bibitem{sidrauski} M.~Sidrauski (1967) Rational choice and patterns of growth in a monetary
economy. American Economic Review, Vol. 57, No. 2, 534-544.
\end{enumerate}
\end{document}